\documentclass{article}

\usepackage{spconf,amsmath,multirow,booktabs}
\usepackage{graphicx}
\usepackage{comment}
\usepackage{xcolor}

\ninept
\title{BW-EDA-EEND: Streaming End-to-End Neural Speaker Diarization \\for a Variable Number of Speakers}

\name{Eunjung Han \quad Chul Lee \quad Andreas Stolcke}
\address{
  Amazon Alexa Speech, Sunnyvale, CA, USA \\
  {\tt \{cehan,chulle,stolcke\}@amazon.com}%
}

\begin{document}

\maketitle

\begin{abstract}
We present a novel online end-to-end neural diarization system, BW-EDA-EEND, that processes data incrementally for a variable number of speakers. The system is based on the Encoder-Decoder-Attractor (EDA) architecture of Horiguchi et al., but utilizes the incremental Transformer encoder, attending only to its left contexts and using block-level recurrence in the hidden states to carry information from block to block, making the algorithm complexity linear in time. We propose two variants: For unlimited-latency BW-EDA-EEND, which processes inputs in linear time, we show only moderate degradation for up to two speakers using a context size of 10 seconds compared to offline EDA-EEND. With more than two speakers, the accuracy gap between online and offline grows, but the algorithm still outperforms a baseline offline clustering diarization system for one to four speakers with unlimited context size, and shows comparable accuracy with context size of 10 seconds. For limited-latency BW-EDA-EEND, which produces diarization outputs block-by-block as audio arrives, we show accuracy comparable to the offline clustering-based system. 
\end{abstract}

\section{Introduction}

Speaker diarization is the process of partitioning an audio stream into homogeneous segments according to speaker identities. Thus, diarization determines ``who spoke when'' in a multi-speaker environment,
with a variety of applications to conversations involving multiple speakers, such as meetings, television shows, medical consultations, or call center conversations. In particular, the speaker boundaries produced by a diarization system can be used to map transcripts generated by a multi-speaker automatic speech recognition (ASR) system into speaker-attributed transcripts \cite{boeddeker2018front, kanda2018hitachi, kanda2019acoustic}. Moreover, speaker embeddings inferred by diarization can help the ASR system adapt to, or focus on the speech of a targeted speaker \cite{kanda2019simultaneous}. 

Conventional speaker diarization systems are based on clustering of speaker embeddings. In this approach, several components are integrated into a single system: speech segments are determined by voice activity detection (VAD); these speech segments are further divided into smaller chunks of fixed size; speaker embeddings are then extracted by speaker embedding extractors for each chunk; finally, those speaker embeddings are clustered to map each segment to a speaker identity \cite{sell2014speaker, garcia2017speaker, li2019discriminative, maciejewski2018characterizing, dimitriadis2017developing, wang2018speaker}. For embeddings, i-vectors \cite{dehak2010front}, x-vectors \cite{snyder2018x}, or d-vectors \cite{wan2018generalized} are commonly used. Clustering methods typically used for speaker diarization are agglomerative hierarchical clustering (AHC) \cite{sell2014speaker, garcia2017speaker, maciejewski2018characterizing}, k-means clustering \cite{dimitriadis2017developing}, and spectral clustering \cite{wang2018speaker}. Recently, neural network-based clustering has been explored \cite{zhang2019fully}. Clustering-based speaker diarization achieves good performance but has several shortcomings. First, it relies on multiple modules (VAD, embedding extractor, etc.) that are trained separately. Therefore, clustering-based systems require careful joint calibration in the building process. Second, systems are not jointly optimized to minimize diarization errors; clustering in particular is an unsupervised process. Finally, clustering does not accommodate overlapping speech naturally, even though recent work has proposed ways to handle regions with simultaneously active speakers in clustering \cite{bullock2020overlap}.

End-to-end neural diarization (EEND) with self-attention \cite{fujita2019end} is one of the approaches that aim to model the joint speech activity of multiple speakers. It integrates voice activity and overlap detection with speaker tracking in end-to-end fashion.  Moreover, it directly minimizes diarization errors and has demonstrated excellent diarization accuracy on two-speaker telephone conversations. However, EEND as originally formulated is limited to a fixed number of speakers because the output dimension of the neural network needs to be prespecified. Several methods have been proposed recently to overcome the limitations of EEND. One approach uses a speaker-wise chain rule to decode a speaker-specific speech activity iteratively conditioned on previously estimated speech activities \cite{fujita2020neural}. Another approach proposes an encoder/decoder-based attractor calculation \cite{horiguchi2020end}. The embeddings of multiple speakers are accumulated over the time course of the audio input, and then disentangled one-by-one, for speaker identity assignment by speech frame.  However, all these state-of-the-art EEND methods only work in an offline manner, which means that the complete recording must be available before diarization output is generated. This makes their application impractical for settings where potentially long multi-speaker recordings need to be processed incrementally (in streaming fashion). 

In this study, we propose a novel method to perform EEND in a blockwise \textit{online} fashion so that speaker identities are tracked with low latency soon after new audio arrives, without much degradation in accuracy compared to the offline system. We utilize the incremental Transformer encoder, where we attend to only its left contexts and ignore its right contexts, thus enabling blockwise online processing. Furthermore, the incremental Transformer encoder uses block-level recurrence in the hidden states to carry over information block by block, reducing computation time while attending to previous blocks. To our knowledge, ours is the first method that uses the incremental Transformer encoder with block-level recurrence to enable online speaker diarization. 

\section{Prior Work}

Horiguchi et al.~\cite{horiguchi2020end} proposed a version of EEND that estimates the number of speakers by aggregating multiple speaker embeddings into a small number of attractors, one per speaker (EEND with Encoder Decoder Attractor, EDA-EEND). To calculate a flexible number of attractors from variable lengths of speaker embedding sequences, they utilize an LSTM-based encoder-decoder \cite{chen2017deep}. Given a $T$-length sequence of $F$-dimensional audio features, $X=[x_1, \dots, x_T], \ x_t \in R^F$, a Transformer encoder computes $D$-dimensional diarization embeddings, $E = [e_1, \dots,  e_T], \ e_t \in R^D$. Then, these embeddings are fed into a unidirectional LSTM encoder, producing the final hidden and cell states ($h_0, c_0$). 
Assuming there are $S$ speakers in $X$, ideally a set of $S$ time-invariant $D$-dimensional attractors, $A = [a_1, \dots, a_S], \ a_s \in R^D$, are then decoded, also using a unidirectional LSTM, with initial states $(h_0, c_0)$ as follows:

\begin{equation*} \label{eq5}
\begin{split}
E &= \text{TransformerEncoder}(X), \\
(h_0, c_0) & = \text{LSTMEncoder}(E), \\
a_s, (h_s, c_s)  &= \text{LSTMDecoder}(0, (h_{s-1},c_{s-1})) \mbox{ for } 1 \le s \le S+1, \\
p_s & = \sigma(\text{Linear}(a_s))\  \mbox{ for } 1 \le s \le S+1, \\
\hat{Y} &= \sigma(E A^T), \ A = [a_{1},\dots,a_{S}]
\end{split}
\end{equation*}
where $0$ is the $D$-dimensional zero vector and used as a constant input for the decoder.
Attractor existence probabilities $p_s$ are estimated using a fully connected layer with a sigmoid function. 
A probability below a chosen threshold indicates that all attractors have been decoded, thereby implicitly determining the number of speakers.
Joint speech activities for $S$ speakers ($\hat{Y}$) are estimated by a matrix multiplication of embeddings and attractors ($EA^T$), followed by a sigmoid function.

\section{Blockwise EDA-EEND}
    \label{sec:blockwise}

\begin{figure}[t]
\centering
\includegraphics[width=0.35\textwidth]{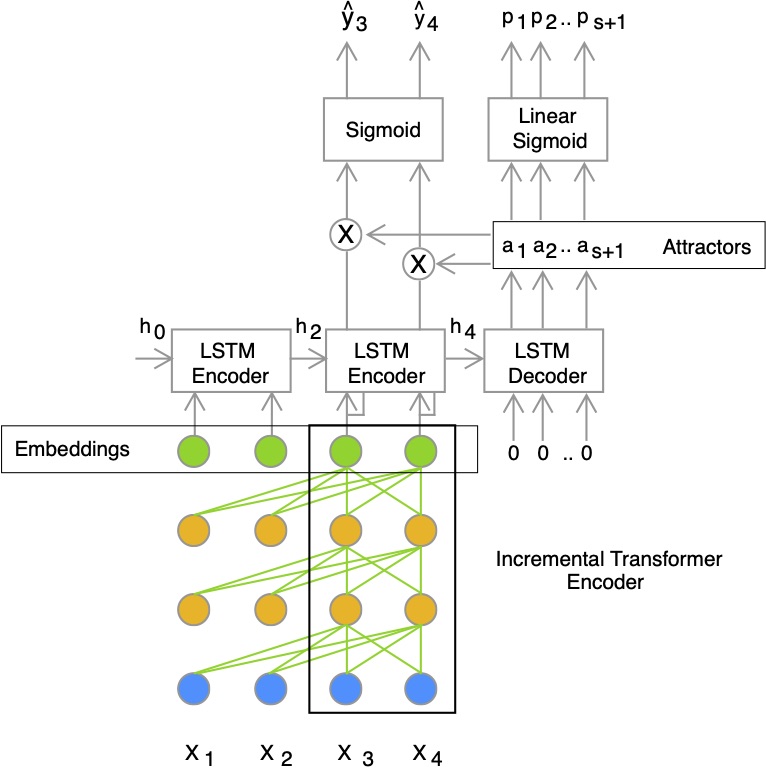}
\caption{Schematic diagram of BW-EDA-EEND}
\label{fig:bw-eda-eend}
\end{figure}

The original EDA-EEND algorithm \cite{horiguchi2020end} performs an \textit{offline} inference operation by first encoding all input frames, aggregating all speaker embeddings, and then finally decoding the attractors. This means that it is not suitable to perform \textit{online} (responsive) speaker diarization. We propose a family of new blockwise EDA-EEND variants that enable online speaker diarization  when (1) latency needs to be limited or
(2) computation needs to be linear in the input length.
To this end, we perform blockwise EDA and EEND operations by incrementally performing various encoding and decoding operations per input audio block, each of only a few seconds in duration.
First, to enable incremental processing, we limit the transformer encoder to attend only to a block's left contexts (ignoring right contexts).
Second, we perform the blockwise recurrence computation throughout the hidden states of the transformer encoder, carrying over information from one block to the next block. This approach was motivated by the segment-level recurrence with state reuse technique used in Transformer XL \cite{dai2019transformer}.
These blockwise recursive operations in the transformer encoder can significantly reduce the overall inference time while also saving memory, since the hidden states from the previous blocks are cached and then reused instead of being re-computed from scratch for every block. More importantly, with a limited context size, the computational complexity of the given Transformer becomes linear-time with respect to audio length, instead of having quadratic complexity.
Although each block can only model local dependency information, the top node of the encoder can still model long-term dependencies via the recurrence.
Also, this is shown to reduce speaker permutations between blocks in our experiments.

Our algorithm is illustrated in Figure~\ref{fig:bw-eda-eend}. Given a $T$-length sequence of $F$-dimensional audio features, $X=[x_1,\ldots,x_T]$, let $X_{b} =[x_{b,1},\cdots,x_{b,W}]$ denote the $b$th block input sequence of size $W$. Given the Transformer encoder with $n$ layers, let $E_b^{i} = [e_{b,1}^{i},\dots,e_{b,W}^{i}]$ denote its hidden states in the $i$th layer.
The Transformer encoder computes the sequence of $D$-dimensional diarization embeddings of size $W$ for the $b$th block, $E_b=[e_{b,1},\ldots,e_{b,W}]$ as:
\begin{equation*} \label{online EEND}
\begin{split}
Q_{b}^{i} &= E_b^{i-1} \mbox{ for } 1 \le i \le n \ (E_b^{0} = X_b), \\
K_{b}^{i} &= V_{b}^{i} = \text{Concat}(E_{b-L:b-1}^{i-1}, E_{b}^{i-1}) \mbox{ for } 1 \le i \le n, \\
E_b^i &= \text{TransformerEncoderLayer}(Q_b^i,K_b^i,V_b^i) \mbox{ for } 1 \le i < n,  \\
E_b &= E_b^n
\end{split}
\end{equation*}
where $E_{b-L:b-1}^{i-1}$ is a sequence of hidden states that are produced from the previous $L$ blocks (i.e., going back $L$ blocks with respect to the current $b$th block) and Concat is a concatenation function of two hidden block sequences along the time dimension.
Motivated by Transformer-XL \cite{dai2019transformer}, both key $K_{b}^{i}$ and value $ V_{b}^{i}$ vectors use the same hidden states from the current block ($E_{b}^{i-1}$) along with other cached hidden states from the previous blocks $E_{b-L}^{i-1}, \ldots, E_{b-1}^{i-1}$. When $L=\infty$, we use all hidden states produced from all previous blocks to $E_{b}^{i-1}$.

Having made the Transformer encoder operation incremental and linear,
we now propose two variants of BW-EDA-EEND that differ in how the attractors are computed.
The first variant, denoted \textcolor{black}{limited-latency BW-EDA-EEND (BW-EDA-EEND-LL)}, computes attractors for each block and produces outputs with limited-latency (according to the block size), suitable for generating outputs online.

The second variant, denoted \textcolor{black}{unlimited-latency BW-EDA-EEND (BW-EDA-EEND-UL)}, computes attractors at the end of the inputs. 
It still has unlimited latency, but limits embedding computation to be linear in input length with a limited context size.

For \textcolor{black}{BW-EDA-EEND-LL}, block dependent $D$-dimensional attractors, $A_b = [a_{b,1},\dots,a_{b,S}]$ are decoded using a blockwise unidirectional LSTM as follows:
\begin{eqnarray*} \label{online EDA1}
\nonumber I_{b} &=& \text{Concat}(E_{b-L:b-1}, E_{b}), \\
\nonumber (h_b, c_b) &= &\text{LSTMEncoder}(I_{b} , (h_{b-L}, c_{b-L})), \\
\nonumber a_{b,s}, (h^*_{b, s}, c^*_{b, s})  &=& \text{LSTMDecoder}(0, (h^*_{b, s-1}, c^*_{b, s-1})) \\
\nonumber && \mbox{ for } 1 \le s \le S+1,\\
\nonumber p_{b,s}  &= &\sigma(\text{Linear}(a_{b,s})) \ \mbox{ for } 1 \le s \le S+1, \\
\nonumber \hat{Y}_b &= & \sigma(E_b A_b^T), \ A_b  =  [a_{b,1},\dots,a_{b,S}]
\end{eqnarray*}
where $(h_b, c_b)$ are the hidden and cell states of the LSTM encoder at the end of the $b$th block (i.e., the LSTM encoder runs unidirectionally along a time window of length $W$-length),
and $E_{b-L:b-1}$ is the sequence of diarization embeddings produced from the previous $L$ blocks.
Note that $(h^*_{b,0}, c^*_{b,0}) = (h_{b}, c_{b})$, meaning that the last hidden state from the LSTM encoder is used as an initial hidden state for the LSTM decoder.
When $L=\infty$, we use $(h_0, c_0)$ instead of $(h_{b-L}, c_{b-L})$ in the LSTM encoder.

For \textcolor{black}{BW-EDA-EEND-UL}, all blockwise LSTM encoding and decoding operations are similar to those of \textcolor{black}{BW-EDA-EEND-LL}, except that block-independent D-dimensional attractors $A = [a_{1},\dots,a_{S}]$ are computed at the end of the last block $m=\lfloor T/W \rfloor$.

During inference, the number of speakers $S$ is not known. 
We estimate $S$, so that the first $S$ attractor existence probabilities are greater or equal to a chosen threshold $\tau$ (typically, $\tau=0.5$), similar to the original EDA-EEND \cite{horiguchi2020end}.
That is, $S = max \{s | s \in Z^{+} \land p_{b,s} \ge \tau \}$.
We use the first $S$ attractors for the subsequent downstream computation in \textcolor{black}{BW-EDA-EEND-UL}.
For the blockwise attractor estimation in \textcolor{black}{BW-EDA-EEND-LL}, we force the number of speakers at each block $S_b$ to be non-decreasing, $S_b \ge S_{b-1}$,
 thereby preventing the model from forgetting the speakers that occurred previously even though they are not active in the current block.
Forcing $S_b$ to be non-decreasing helps all attractors to be kept in memory for possible matching when the corresponding speakers reappear later.

Since we re-decode all attractors after each block, we need to ensure their consistency in both order and values. We use several heuristics in the inference phase to encourage consistency of attractors, and therefore speaker labels, across blocks. In ablation experiments (Table~\ref{tab:DER reduction for online}), we show that the individual heuristics incrementally improve the overall performance of our BW-EDA-EEND algorithms. 
First, we reorder (permute) attractors (in a greedy search) so that the cosine similarities between previous and current block are maximized.
Additionally, we average the values of attractors from the previous block and the current block to make speaker representations vary more smoothly as we process more audio information.

Finally, we allow the shuffling of diarization embeddings (outputs of the Transformer encoder, used as inputs to the LSTM encoder) \textit{across} blocks. 
The use of LSTMs as a Markovian memory mechanism and the shuffling heuristic comes from the original EDA algorithm. The purpose is to incrementally add speaker embeddings to the memory representation. Shuffling enforces invariance of the memory mechanism with respect to the order in which speakers appear in the input.
We shuffle diarization embeddings randomly within the current and $L_s$ context blocks before sending them to the LSTM encoder. We set $L_s = L$ if $L$ is finite, and $L_s = 0$ if $L=\infty$.
\section{Experiments}

For EDA-EEND, we used the simulation recipe in \cite{fujita2019end} to generate meeting mixtures of one to four speakers. We simulated meeting mixtures based on Switchboard-2 (Phase I, II, III), Switchboard Cellular (Part 1 ,2), and the 2004-2008 NIST Speaker Recognition Evaluation (SRE) datasets. All recordings are telephone speech sampled at 8 kHz. Since there are no time annotations in these corpora, we extracted speech segments using speech activity detection (SAD) on the basis of a time-delay neural network and statistics pooling from a Kaldi recipe. We added noises from 37 background noise recordings from the MUSAN corpus \cite{snyder2015musan}. We used a set of 10,000 room impulse responses (RIRs) from the Simulated Room Impulse Response Database \cite{ko2017study}. We also prepared real conversations from the CALLHOME corpus \cite{callhome}. We divided the CALLHOME data into two subsets: an adaptation set of 250 recordings and a test set of 250 recordings, randomly split according to the recipe in \cite{fujita2019end}. 

The input audio features are 23-dimensional log-scaled Mel filterbanks with a 25-ms frame length and 10-ms frame shift. Each feature vector is concatenated with both the previous and following seven frames (for a total of 15 frames).
We then subsample the concatenated features by a factor of ten.
Consequently, $(23 \times 15)$-dimensional input features sampled every 100 ms are fed into the Transformer encoder block for all EDA-EEND systems.

We used four stacked Transformer encoders with 256 attention units, containing four heads each. We first trained all three systems using simulation data with two speakers for 100 epochs, using the Adam optimizer with a learning rate schedule with 100,000 warm-up steps.
We then finetuned the two-speaker models with simulated data with one to four speakers, for 100 epochs.
Finally, we finetuned the models using CALLHOME adaptation data (250 recordings with two to seven speakers per call) and evaluated performance on CALLHOME test data (with two to six speakers per call).
In Tables~\ref{tab:DER reduction for online} and~\ref{tab:DER Simulated and CH}, we show results separately for subsets with different numbers of speakers (148 recordings with 2 speakers, 74 recordings with speakers, 20 recordings with 4 speakers).

We evaluated all systems by their diarization error rates (DER).
As is standard, a tolerance of 250\,ms when comparing hypothesized to reference speaker boundaries was used.

\section{Results}

\begin{table}[tb]
	\caption{DER (\%) of offline and blockwise online diarization methods on CALLHOME two-speaker data.
	Block size $W = 10s$ and context size $L = \infty$.
	Variables controlling attractor computation: by {\bf Block} (versus per utterance),
	    {\bf Reorder} (to match previous block),
	    {\bf Average} (across blocks),
	    {\bf Shuffle} (embeddings across blocks).}
	\label{tab:DER reduction for online}
	\centering
\resizebox{\linewidth}{!}{
\begin{tabular}{ c c c  c c c c r}
	\toprule
	& \bf Train &  \bf Inference & \multicolumn{4}{c}{\bf Attractor computation} & \bf DER \\
	\cmidrule{4-7}
	& & & \bf Block & \bf Reorder & \bf Average & \bf Shuffle & \\ 
	\midrule
	1 & offline & offline       & no & & & & 9.02 \\
	\midrule
    
	2 & offline  & causal & no &   &  &  & 12.73 \\
	3 & causal   & causal & no &   &  &  & 10.05 \\
	\midrule
	4 & offline  & causal & yes & no  & no & no & 24.24 \\ 
	5 & causal   & causal & yes & no  & no & no & 18.07  \\ 
	6 & causal   & causal & yes & yes  & no & no &  14.26  \\	
	7 & causal   & causal & yes & yes  & yes & no &  12.91  \\
	8 & causal   & causal & yes & yes  & yes & yes &  11.82 \\
	\bottomrule
\end{tabular}
}
\end{table}

\begin{table*}[t]
	\caption{DERs (\%) of offline and online diarization methods on simulated and CALLHOME data. Speaker number is estimated by the algorithm. As a baseline, the x-vector clustering result is included as reported in \cite{fujita2019end, fujita2020neural, horiguchi2020end}. We use BW-EDA-EEND-UL when attractors are computed at the end of the utterances and BW-EDA-EEND-LL when attractors are computed block by block.}
	\label{tab:DER Simulated and CH}
	\centering
	\footnotesize
	\begin{tabular}{ l  c c | r  r  r  r | r r r }
	\toprule
	        & & & \multicolumn{7}{c}{\bf Number of speakers} \\
	\cmidrule{4-10} 
	\bf Model & \bf Attractor & \bf Embedding & \bf 1 & \bf 2 & \bf 3  & \bf 4 & \bf 2 & \bf 3 & \bf 4 \\
	       & \bf computation & \bf shuffling  & \multicolumn{4}{c|}{\bf Simulated} & \multicolumn{3}{c}{\bf CALLHOME} \\
	\midrule
	Offline x-vector  &  &  & 37.42  & 7.74 & 11.46 & 22.45      & 15.45 & 18.01 & 22.68 \\
	Offline EDA-EEND     & by utterance  & within utterance & 0.27  & 4.18 & 9.66 & 14.21         & 9.02  & 13.78 & 20.69  \\
	\midrule
	Online $W=10, L=\infty$ & by utterance & within block & \bf{0.28} & \bf{4.22} & \bf{11.17} & \bf{21.04}   & \bf{10.05} & \bf{16.59} & 25.50 \\ 
	Online $W=10, L=1$      & by utterance & within block & \bf{0.30} & \bf{4.42} & 13.35 & 22.71             & \bf{11.73} & 19.87 & 28.03 \\ 
    \midrule
	Online $W=10, L=\infty$ & by block & within block  & \bf{1.72} & 8.46 & 20.60 & 37.17   & \bf{12.91} & 22.04 & 30.62 \\
    Online $W=10, L=1$      & by block & within block  & \bf{1.85} & 10.99 & 22.23 & 36.72  & 16.84 & 26.01 & 28.91 \\ 
    Online $W=10, L=\infty$ & by block & across blocks & \bf{1.03} & \bf{6.10} & 12.58 &  \bf{19.17}   & \bf{11.82} & 18.30 & 25.93 \\ 
    Online $W=10, L=1$      & by block & across blocks & \bf{2.49} & \bf{7.53} & 16.65 & 24.50   & 16.18 & 19.35 & 27.52 \\ 
    
	\bottomrule
    \end{tabular}
\end{table*}

Table~\ref{tab:DER reduction for online} shows the effect of various features of the \textcolor{black}{BW-EDA-EEND} algorithms. 
As described in Section~~\ref{sec:blockwise}, we trained the Transformer encoder to only attend to its left contexts and reuse the hidden states from the previous blocks (Train = causal) instead of using entire utterances to compute attention weights (Train = offline).
During inference, we reuse the hidden states from the previous blocks (Inference = causal) or recompute them from scratch for every block (Inference = offline).
As shown in Table~\ref{tab:DER reduction for online}, incremental training of the Transformer encoder improved accuracy since training and inference mechanisms now are consistent.
For  \textcolor{black}{unlimited-latency BW-EDA-EEND}, accuracy improved by 2.7\% absolute (row~2 vs.~3 in Table~\ref{tab:DER reduction for online}) and for  \textcolor{black}{limited-latency BW-EDA-EEND} accuracy improved by 6.2\% absolute (row~4 vs.~5), when the algorithms are evaluated for two-speaker CALLHOME data using a block size of 10 seconds.

Next, during attractor inference in  \textcolor{black}{BW-EDA-EEND}, we compare several heuristics to improve accuracy (rows 4-8 in Table~\ref{tab:DER reduction for online}).
By reordering attractors, accuracy is improved by 3.8\% absolute (row~5 vs.~6).
Through attractor averaging, accuracy improves by 1.3\% absolute (row~6 vs.~7). By shuffling embeddings across blocks, accuracy improves by 1.1\% absolute (row~7 vs.~8).

We evaluated the proposed BW-EDA-EEND on simulation data with a variable number of speakers.
Note that we trained the initial model with simulated mixtures of two speakers, then finetuned it with more simulation data, now mixing up to four speakers.
The results for various context sizes are shown in Table~\ref{tab:DER Simulated and CH} (left half). We use the algorithm of Table~\ref{tab:DER reduction for online}, row~3 to report accuracy for  BW-EDA-EEND-UL (attractors are computed at ends of utterances) and the algorithm in Table~\ref{tab:DER reduction for online}, rows~7 and~8, for  BW-EDA-EEND-LL (attractors are computed block-by-block).
First, when attractors are computed at the ends of the utterances, accuracy degrades only moderately up to two speakers, using either unlimited or 10 seconds ($L=1$) context.
With more than two speakers in a conversation, the accuracy gap between online and offline increases, but \textcolor{black}{BW-EDA-EEND-UL} with unlimited left context still outperforms the baseline clustering-based system \cite{fujita2019end, fujita2020neural, horiguchi2020end} for one to four speakers, or shows comparable accuracy with a single block of context.

\begin{figure}[tb]
\centering
\includegraphics[width=0.5\textwidth, height=0.3in]{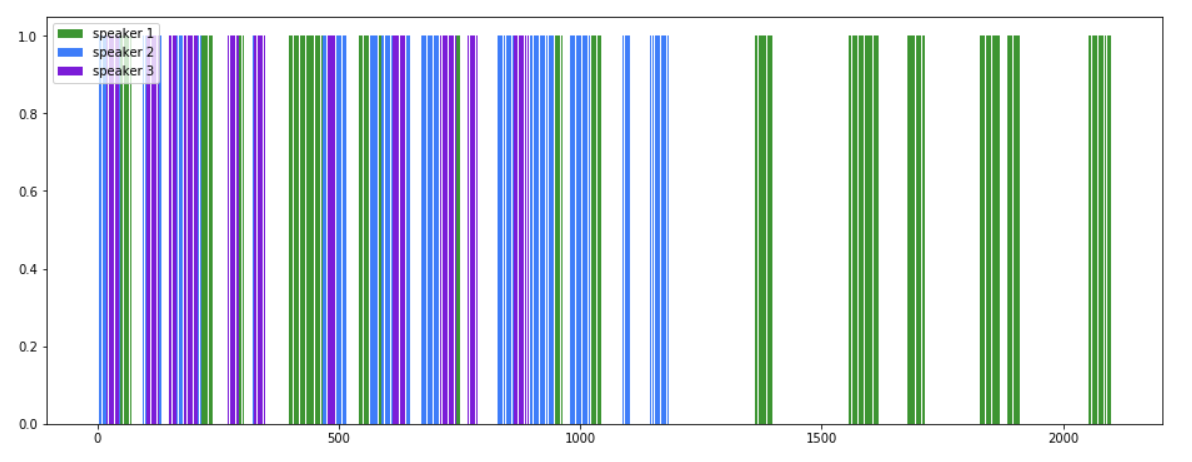}
\includegraphics[width=0.5\textwidth, height=0.3in]{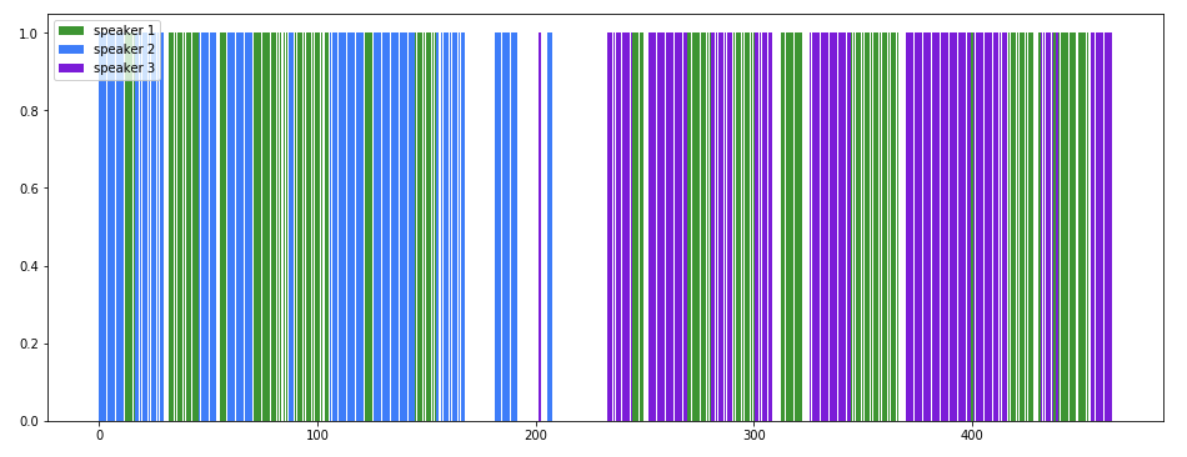}
\caption{Visualization of speaker turns for simulated (top) and CALLHOME data (bottom). There are three speakers in these examples.}
\label{data_n3}
\end{figure}

When attractors are computed block-by-block, \textcolor{black}{BW-EDA-EEND-LL} does not perform well unless diarization embeddings are shuffled \textit{across} blocks. This shows the importance of shuffling embeddings for the LSTM to learn order-invariance during encoding.
When frame-level embeddings are shuffled across blocks, we see comparable accuracy to a offline clustering-based (x-vector) system with unlimited left context or a single block of context.
We conclude that it is crucial that EDA receives embeddings from \textit{all} the speakers to accurately compute attractors for all the speakers who have spoken so far.
If only a subset of speakers are active in a given block, it seems that information about the previous speakers tends to be forgotten. In simulation data, we often observe only a subset of speakers active in the tail of conversation, hurting the accuracy when attractors are computed block-by-block (Fig.~\ref{data_n3}, top half).

Furthermore, we evaluated \textcolor{black}{BW-EDA-EEND} on real conversation data. For testing on real recordings, we finetuned the model first trained with simulation data, using the CALLHOME adaptation set.
The results for various block sizes are shown in Table~\ref{tab:DER Simulated and CH} (right half).
Consistent with previous findings, \textcolor{black}{BW-EDA-EEND-UL} produces similar accuracy as that of offline EDA-EEND for up to two speakers when context is unlimited or 10 seconds.
When attractors are computed block-by-block, \textcolor{black}{BW-EDA-EEND-LL} does not perform well if diarization embeddings are shuffled within blocks,
but results in accuracy comparable to the clustering-based system when frame-level embeddings are shuffled across blocks, for up to three speakers and left context unlimited or 10 seconds ($L=1$) context.
Similar to simulation data, we often find that a subset of speakers is active only in part of a conversation, hurting the accuracy when attractors are computed block-by-block (Fig.~\ref{data_n3}, bottom half).

\section{Conclusions}

We implemented two versions of a blockwise online variant of EDA-EEND, BW-EDA-EEND-UL (unlimited latency) and BW-EDA-EEND-LL (limited latency).
Blockwise online processing is enabled by utilizing an incremental Transformer encoder that attends to only its left contexts and ignores its right contexts and uses block-level recurrence in the hidden states to carry over information between blocks, which makes algorithm complexity linear in time. BW-EDA-EEND-UL shows only moderate degradation of accuracy for up to two speakers using either unlimited or 10-second context, compared to offline EDA-EEND.
BW-EDA-EEND-LL has accuracy comparable to an offline clustering-based system when frame-level embeddings are shuffled across blocks. Future algorithmic improvements should address the consistency of attractor direction and ordering over time, as blocks are processed incrementally.

We observe that multi-speaker training data as simulated in prior work \cite{fujita2019end} (and adopted here for compatibility) is not very realistic when compared to real conversational data, in terms of turn-taking behavior (as shown in Fig.~\ref{data_n3}),
and suspect that this may limit the effectiveness of model training. In future work, we plan to modify the simulation algorithm to create more realistic meeting mixtures by adopting the recently proposed method that was used to create LibriCSS test data \cite{chen2020libricss}.
We also believe that test data for EEND needs to become more realistic,
moving from mixtures of telephone speech channels (CALLHOME) to far-field recordings of multiple speakers speaking and interacting in the same room.

\section{Acknowledgments}
We would like to thank our colleague Sundararajan Srinivasan, as well as Shinji Watanabe and other members of
the 2020 JSALT workshop on ``Speech Recognition and Diarization for Unsegmented Multi-talker Recordings with Speaker Overlaps'' for help with data preparation and valuable discussions.

\normalsize

\bibliographystyle{IEEEtran}
\bibliography{mybib}

\end{document}